\documentclass[showpacs,amsmath,showkeys,twocolumn]{revtex4}

\usepackage[latin1]{inputenc}
\usepackage{graphicx}
\usepackage{dcolumn}
\usepackage{bm}

\newcommand{\he}[1]{$^#1$He}

\begin{document}


\title{Phase boundary detection for dilution refrigerators}

\author{E. ter Haar}
 \email{ewout@if.usp.br}
\author{R. V. Martin}

\affiliation{%
DFMT, Instituto de F\'{i}sica, Universidade de S\~{a}o Paulo, C.P.
66.318, 05315-970 S\~{a}o Paulo, SP, Brazil}

\date{\today}

\begin{abstract}
We describe a device to conveniently measure the positions of the
phase boundaries in a dilution refrigerator. We show how a simple
modification of a standard capacitive level gauge (segmentation of
one of the electrodes) permits a direct calibration of the
capacitance versus phase boundary position. We compare this direct
calibration with the indirect procedure that must be adopted for a
conventional capacitive level gauge. The device facilitates the
correct adjustment of the \he{3}/\he{4} fraction in the dilution
refrigerator.
\end{abstract}

\pacs{07.20.Mc,}

\keywords{Dilution Refrigerators, Phase-boundary detection}

\maketitle

\section{Background}
Dilution refrigerators (DR) have made the temperature region below
0.5 K widely available to the research community. While the
continuing development of the technologies involved have made
these machines very reliable, good diagnostic devices are
essential for their continuing and reliable operation.

Various problems can impede the correct operation of a dilution
refrigerator. Leaks to vacuum, blockages and thermal shorts
usually lead to catastrophic failure, and are easy to diagnose,
but difficult to solve. In contrast, not having the right amounts
of \he{3} and \he{4} in the system is easily solved, but may not
be readily recognized.

For optimal performance of the DR, the total amount of mixture,
and the \he{3}/\he{4} fraction of the mixture must be {\em tuned},
which means putting the phase boundaries in the still (between the
diluted liquid and the vapour phase) and in the mixing chamber
(between the concentrated \he{3} and the \he{3} diluted in \he{4}
phases) at the correct positions. The still level is determined by
the total quantity of mixture in the DR. If this level is too low
(below the still heater or in the heat exchanger) it will be
difficult to pump and circulate the \he{3}. If the level is too
high (in the pumping line of the still) again it will be difficult
to pump the \he{3} and large amounts of \he{4} may be circulated.
Both circumstances will degrade the performance of the DR. At the
phase boundary in the mixing chamber level the dilution cooling
takes places and it is the coldest position in the mixing chamber.
This phase boundary must not be up in the heat exchangers, since
the cooling power of the DR will suffer. It may be advantageous in
some cases to be able to control the position of the phase
boundary to put it close to the sample.

Tuning the amounts of \he{3} and \he{4} is not a one time
operation because changes may occur over time, leading to degraded
performance. If the sample is mounted in the mixing chamber, the
mixing chamber volume will depend on the sample size, which will
lead to a different optimal quantities of \he{3} and \he{4}. For
small DR´s with large \he{3} circulation pumps (to get high flow
rates and cooling powers) the large fraction of dead volume behind
the pumps makes the \he{3}/\he{4} fraction in the DR depend on the
operating conditions. If the circulation rate, and therefore the
inlet pressure of the \he{3} return line, is varied the amount of
\he{3} in the DR will change too.

For all of these reasons, a good measurement of the positions of
the phase boundaries in the still and in the mixing chamber is
very useful for finding optimal working conditions of the DR.
Although it is possible to infer the positions of the phase
boundaries using only thermometers at strategic places
\cite{richardson98}, it is preferable to have real level gauges in
the still and the mixing chamber, providing a direct measurement.

\section{Phase boundary level gauges}

We focus here on capacitive level gauges \cite{celik2001} which
may be parallel plates or concentric cylinders. The principle of
operation is simplicity itself: the capacitance is given by
\begin{equation}
C(x) = C_{s} + C(0)[\epsilon_{a} + x(\epsilon_{b}-\epsilon_{a})]
\label{eqn1}
\end{equation}
with $x$ the fractional level, $\epsilon_{a,b}$ the relative
dielectric constants of the phases above and below the phase
boundary at $x$, and $C_{s}$ the inevitable stray capacitance.
Because of this stray capacitance, the most convenient mode of
operation is to measure simply $C(0)$ and $C(1)$, the `empty'
($x=0$) and `full' ($x=1$) capacitance readings. After these
measurement the device is fully calibrated: the sensitivity of the
device, $dC/dx = C(1) - C(0)$ and a fixed point (e.g. $C(0)$) are
known.

In the case of phase-boundary detection in dilution refrigerators,
it is difficult to perform this simple calibration under operating
conditions. We may not be willing to fill up the still, since it
is difficult to remove the excess \he{4} later. Usually it is also
not feasible to add enough \he{3} to the mixture so that the phase
boundary lowers enough to cover the whole level gauge in the
mixing chamber.

One can then resort to an indirect calibration using phases with a
known dielectric constant and Eq.~\ref{eqn1} (one can use \he{4}
and vacuum in the still or \he{3}/\he{4} mixture and vacuum in the
mixing chamber). But the dielectric constants are a function of
temperature (through the density), which must be accurately
measured during the calibration. The stray capacitance may not be
constant over the course of the experiment; it may vary with the
level of the helium bath or with the temperature of the level
gauge itself. The electrodes may not be exactly parallel,
invalidating Eq.~\ref{eqn1}. In the mixing chamber small
calibration errors are exacerbated because the difference between
the dielectric constants of the phases is five times less than in
the still.

\begin{figure}[tb]
  \includegraphics[width=5cm]{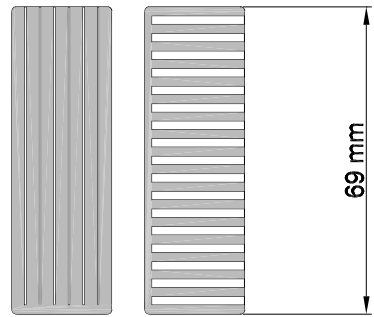}
  \caption{The two electrodes of the capacitive level gauge.
  One of the electrodes is segmented, leading to capacitance steps
  as the phase boundary moves.}
  \label{fig1}
\end{figure}

These kind of difficulties may be the reason that despite being
extremely useful diagnostic devices, capacitive phase boundary
detectors are not widely used, especially in the mixing chamber.
To solve the calibration problem we have used a segmented
capacitor as shown in Figure~\ref{fig1}. It is constructed from
two pieces of standard circuit board separated by $\approx
0.5$~mm. One of the electrodes is segmented horizontally, by
etching away a number of 4~mm wide ``fingers" leaving only a thin
strip of copper at the edge for electric contact. In effect, we
have a stack of small capacitors, connected in parallel. When a
phase boundary moves up or downwards in the device, the
capacitance will change in steps. A calibration of the device can
now be done by moving the phase boundary only one or two segments
upward or downwards. The sensitivity $dC/dx$ around the position
where the phase boundary was varied follows immediately. The
advantage of this scheme is that it is a direct calibration of the
sensitivity, which does not involve knowledge of the dielectric
constants. In the still a segmented capacitor is especially
convenient. As soon as the first segment is covered during the
condensation, as indicated by the first plateau in the capacitance
reading, the level gauge is calibrated and ready for use.

For a mixing chamber phase boundary level gauge, the phase
boundary between the concentrated \he{3} phase and the diluted
phase in the mixing chamber can be varied by adjusting the amount
of \he{3} in the DR. One can perform a so-called `one-shot',
achieved by temporarily closing off the returning \he{3}, while
continuing to pump \he{3} out of the still. In this situation the
amount of \he{3} in the DR will diminish, and the phase boundary
in the mixing chamber will move upwards. On the other hand,
condensing more \he{3} will move the phase boundary downwards. For
the calibration to succeed, it is necessary to move the phase
boundary monotonically up or downwards, but this is not difficult
to achieve.

\section{Results}

\begin{figure}[tb]
  \includegraphics[width=8.5cm]{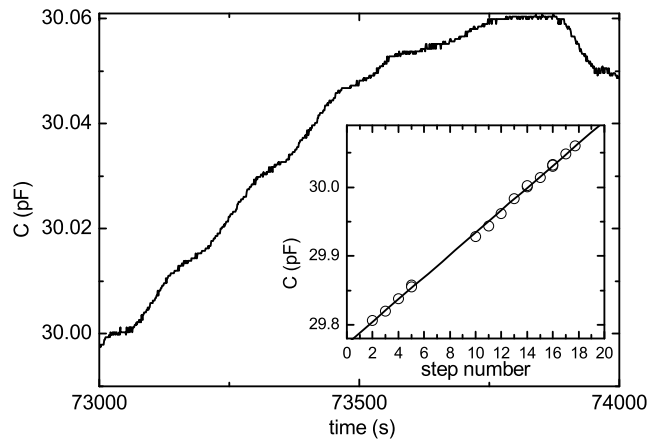}
  \caption{Capacitance of the level gauge during removal of
  \he{3} out of the dilution refrigerator. Inset: the step
  capacitances were recorded for a range of phase boundary positions.
  A linear fit gives a calibration of 0.0162~pF/segment
  ($\approx 0.0042$~pF/mm).}
  \label{fig2}
\end{figure}

We have performed an experiment with the level gauge depicted in
Figure~\ref{fig1} in the mixing chamber of one of our plastic
dilution refrigerators. In addition to the present device, there
were four vibrating wire viscometers (horizontal stretched wires)
at known positions in the region of the phase boundary. For the
purposes of this experiment they served as additional (discrete)
level indicators of the phase boundary, because the viscosity of
pure \he{3} is much larger than that of the diluted phase.
Figure~\ref{fig2} shows the result of one particular `one-shot'.
The capacitance steps are clearly resolved, allowing us to infer
the calibration constant. Controlling the phase boundary over
almost the whole length of the mixing chamber allowed us to
observe almost all steps (see inset in Figure~\ref{fig2}). A
linear fit gives the sensitivity of our device (0.0162~pF/segment,
$\approx 0.0042$~pF/mm, $dC/dx = 0.287$~pF). The (expected)
linearity shows that our electrodes were indeed parallel.

To know the absolute position of the phase boundary, it is still
necessary to measure a capacitance value at one known position of
the interface. The `empty' value is a convenient point (only
diluted phase in the mixing chamber; the phase boundary is above
the level gauge). In our experiment, we also used the large change
of the resonance width of our horizontal viscometers when the
phase boundary passed them. Both methods lead to an `empty' value
$C(1)=30.061$~pF. In the case of a still level gauge the most
convenient point would be the empty value $C(0)$ (only vapour
phase in the gauge).

After the calibration, we were able to monitor and control the
phase boundary in the mixing chamber to within a fraction of a
segment (a segment corresponded to 3.9~mm). This knowledge greatly
helped us to operate the DR with different flow rates (from 100 to
1000~$\mu$mol/s) because the dead volume behind our large \he{3}
circulation pump was a considerable fraction of the total volume
of the DR, leading to large variations of the phase boundary in
the mixing chamber as a function of the flow rate.

The segmented capacitor gives us an opportunity to compare its
direct calibration with the procedure which has to be adopted for
a conventional non-segmented capacitor. If it is not feasible to
measure `full' and `empty' values under operating conditions, one
can calibrate the gauge with phases of known dielectric constants.
We calculated the dielectric constants of the various helium
phases using the Clausius-Mossotti relation with the molar volumes
found in reference~\cite{dewaele92}

We measured the value of the level gauge in vacuum, $C_{v} =
28.832$~pF, and the value with only diluted \he{3} phase, $C(1) =
30.061$~pF. The sensitivity for the \he{3} -- \he{3}/\he{4} phase
boundary is then

\begin{eqnarray*}
dC/dx  = && (C(1) - C_{v}) \frac{\epsilon_{34} - \epsilon_{3}}{\epsilon_{34} -1} = \\
          && (C(1) - C_{v}) \cdot 0.238 = 0.293~\text{pF}
\end{eqnarray*}
where $\epsilon_{34} = 1.0562$ and $\epsilon_{3} = 1.0428$ for $T
\leq 100$~mK. This result must be compared to a directly measured
value of $0.287$~pF. Thus it turns out that this indirect
calibration, is actually quite good in the case of our level
gauge. We note that for a calibration of a conventional level
gauge in the still there are no alternatives, short of filling the
whole length of the capacitor.

The level gauge can also be calibrated by pure \he{4} at a known
temperature (density). In practice, this is not as easy as it
sounds, because at the beginning of a DR experiment we may be
unwilling to fill the DR with (difficult to remove) pure \he{4}.
Nevertheless, it is interesting to see how accurate such a
calibration is. Note that the dielectric constant of \he{4}
depends strongly on temperature (through the density) above $T
\approx 1.5$~K. We used \he{4} at $T=1.17$~K (measured with a
calibrated Ge thermometer, $\epsilon_{4} = 1.0573$), and vacuum as
the calibration phases. The result was $C(0) = 28.832$~pF and
$C(1) = 30.106$~pF. The scaled sensitivity for the phase boundary
in the mixing chamber is $(C(1)-C(0))\cdot 0.234 =
0.298~\text{pF}$, which must be compared to the 0.287~pF measured
directly.

Our conclusion is that a conventional non-segmented capacitor can
be used as a level-gauge, if careful measurements of the `empty'
(vacuum) and `full' (with a phase of known dielectric constant)
capacitance values are made. The use of a segmented capacitive
phase boundary level gauge is more convenient. It makes a direct
calibration of $dC/dx$ in units of pF/segment possible under
operating conditions. This calibration does not involve knowledge
of the dielectric constants, and can be done by moving the phase
boundary only one or two segments. In addition, the level gauge
can be checked for non-linearities.

\section{Acknowledgment}
This research was partly supported by the Brazilian agency
Fundação de Amparo à Pesquisa do Estado de São Paulo

\section{Copyright}
Copyright 2004 American Institute of Physics. This article may be
downloaded for personal use only. Any other use requires prior
permission of the author and the American Institute of Physics.
The following article appeared in  Review of Scientific
Instruments -- September 2004 -- Volume 75, Issue 9, pp. 3071-3073
and may be found at \url{http://link.aip.org/link/?rsi/75/3071}

\bibliography{levelgauge}

\pagebreak

%
%

%
%
%
%
%

\end{document}